\documentclass[11pt,nofootinbib,floatfix,onecolumn]{revtex4}

\usepackage{amsmath,ulem}
\usepackage{amssymb}
\usepackage{graphicx}
\usepackage{rotating}
\usepackage{color}
\usepackage{multirow} 
\usepackage[T1]{fontenc} 
\usepackage{slashed}
\usepackage[mathscr]{eucal}
\usepackage{color}
\usepackage{longtable}
\usepackage{ulem}

\usepackage{pifont} 
\newcommand{\tickYes}{\checkmark} 
\newcommand{\tickNo}{\hspace{1pt}\ding{55}} 

\newcommand{\MET}{$\slashed{E}_T$}
\newcommand{\EQ}{Eq.}

\newcommand{\TABLE}{Table}

\newcommand{\AddrBonn}{%
Bethe Center for Theoretical Physics \& Physikalisches Institut der 
Universit\"at Bonn, \\
Nu{\ss}allee 12, 53115 Bonn, Germany
}

\newcommand{\AddrOrsay}{%
Laboratoire de Physique Th\'eorique, CNRS -- UMR 8627, Universit\'e de Paris-Sud 11\\ F-91405 Orsay Cedex, France}

\preprint{BONN-TH-2012-31,LPT-12-110}
\begin{document}

\title{General NMSSM signatures at the LHC}

\author{H. K. Dreiner}\email{dreiner@uni-bonn.de}
\author{F. Staub}\email{fnstaub@physik.uni-bonn.de}
\affiliation{\AddrBonn}

\author{A. Vicente} \email{avelino.vicente@th.u-psud.fr}
\affiliation{\AddrOrsay}

\begin{abstract}
  We study the possible LHC collider signatures in the next-to-minimal
  supersymmetric Standard Model (NMSSM). The general NMSSM consists of
  29 supersymmetric (SUSY) particles which can be mass ordered in
  $29!\simeq 9\cdot10^{30}$ ways. To reduce the number of hierarchies
  to a more manageable amount we assume a degeneracy of the sfermions
  of the first two generations with the same quantum numbers. Further
  assumptions about the neutralino and chargino masses leave 15
  unrelated parameters.  We check all $15!\approx 10^{12}$ relevant
  mass orderings for the dominant decay chains and the corresponding
  collider signatures at the LHC. As preferred signatures, we consider
  charged leptons, missing transverse momentum, jets, and $W,\,Z$ or
  Higgs bosons.  We present the results for three different choices of
  the singlet to Higgs coupling $\lambda$: (a) small: $O(\lambda)<
  O(Y_{\tau})$, (b) large: $O(\lambda)\simeq O(Y_{top},Y_b, Y_\tau)$
  and (c) dominant: $O(\lambda)>O(Y_{top})$. We compare these three
  scenarios with the MSSM expectations as well as among each other. We
  also mention a possible mass hierarchy leading to 7 jets plus 1
  lepton signatures at the LHC and comment briefly on the consequence
  of possible $R$-parity violation.
\end{abstract}

\maketitle

\section{Introduction} 
There is strong evidence that a particle similar to the Standard Model Higgs boson exists with a mass in the range between 122 and 128~GeV \cite{Atlas:2012gk,CMS:2012gu}.  Assuming this is the Higgs boson, this has significant implications for supersymmetric extensions of the Standard Model (SM) capable of ameliorating the (little) hierarchy problem \cite{Gildener:1976ai}.  In particular, in the Constrained Minimal Supersymmetric Standard Model (CMSSM) the fine tuning needed to achieve this Higgs mass is large, requiring a cancellation between (in the CMSSM) uncorrelated parameters of order 1 part in 300~\cite{Ghosh:2012dh,Balazs:2012qc,Cassel:2011tg,Bechtle:2012zk}.  The overall fit of the CMSSM to the low-energy and LHC data, including the Higgs, is also poor \cite{Bechtle:2012zk,Buchmueller:2012hv,Fowlie:2012im,other}.  In the context of the more general MSSM, allowing for non-universal supersymmetry breaking parameters, defined close to the electroweak scale it is easier to find valid regions of parameter space to explain the Higgs mass.  However, still large stop masses and A-terms are needed~\cite{Heinemeyer:2011aa}.  This tension gets significantly reduced if a new gauge singlet is added to the particle content which has a superpotential coupling to the MSSM doublet Higgs fields. The easiest example of this kind of model is the next-to-minimal, supersymmetric Standard Model (NMSSM) \cite{Fayet:1974pd,Nilles:1982dy,Ellis:1988er,Ellwanger:2009dp}.  New F-Term contributions already at tree-level help to increase the Higgs mass which is bounded in the MSSM (at tree-level) to $m_h \leq M_Z$. This makes it much easier to obtain Higgs masses in the preferred mass range in much larger areas of parameter space and thus reduces the fine-tuning \cite{Ross:2012nr}.

However, the new singlet state might not only have an impact on the mass of the light Higgs boson but also on the collider phenomenology: for instance, an additional light, singlino-like neutralino can appear in the SUSY cascade decays.  Therefore, we shall compare all possible signatures of the general NMSSM with those possible in the general MSSM. Previously, the general signatures of the MSSM based on 9 and 14 free mass parameters at the SUSY scale have been studied in \cite{Konar:2010bi} and \cite{Dreiner:2012wm}, respectively. In the latter case the third generation was treated separately.  We take the scenario with 14 mass parameters and add a 15th parameter, the singlino mass. This leads in general to $15! \sim 1.3\cdot10^{12}$ mass hierarchies. We categorize all signatures by the number of charged leptons, jets, massive bosons and the presence or absence of missing transverse energy ($\slashed{E}_T$, which is actually missing transverse momentum, $\slashed{p}_T$). In this context we study three different NMSSM scenarios: dominant $\lambda$, large $\lambda$ and small $\lambda$, where $\lambda$ denotes the singlet Higgs coupling, \textit{cf.} Eq.~(\ref{superpot}). In the first case, $\lambda$ is even larger than the top Yukawa coupling, while in the second case it is comparable to the size of the third generation Yukawa couplings. In the third case it is not larger than a second generation Yukawa coupling and we are in the MSSM limit.

The rest of the paper is organized as follows: In sec.~\ref{sec:model}
we give the basic definitions and conventions used throughout the
paper. We explain in detail our approach and the underlying
assumptions in sec.~\ref{sec:analysis}. In sec.~\ref{sec:results} we
discuss our results before we conclude in sec.~\ref{sec:conclusion}.

\section{Model definitions}
\label{sec:model}
In the following we consider the next-to-minimal supersymmetric standard model
(NMSSM). For a detailed introduction to the NMSSM we refer the interested reader to
Ref.~\cite{Ellwanger:2009dp}. In the NMSSM the particle content of the MSSM is extended 
by one chiral superfield which is a gauge singlet: $\hat{S}\, ({\bf
1},{\bf 1},0)$. In parentheses we give the SM gauge quantum numbers
with respect to $SU(3)_c \times SU(2)_L \times U(1)_Y$.
The other chiral superfields of the supersymmetric SM read
$\hat{q}_a\, ({\bf 3},{\bf2 },\frac{1}{6})$, $\hat{\ell}_a\,({\bf
1},{\bf 2},-\frac{1}{2})$, $\hat {H}_d\, ({\bf 1},{\bf
2},-\frac{1}{2})$, $\hat{H}_u\,({\bf1},{\bf 2},
\frac{1}{2})$, $\hat{d}^c_a\,({\bf \bar{3}},{\bf1},\frac{1}{3})$, $
\hat{u}^c_a\,({\bf\bar{3}},{\bf 1},\frac{2}{3})$, $\hat{e}^c_a\, ({\bf
1},{\bf 1},1)$, where $a=1,2,3$ is a generation index. The vector superfields 
are the same as in the MSSM: $\tilde{g}_\alpha \, ({\bf
8},{\bf 1},0)$, $\tilde{W}^i\, ({\bf 1},{\bf 3},0)$, $\tilde{B}\,({\bf
1},{\bf 1},0)$.   Using these superfields and demanding an additional
$Z_3$ symmetry\footnote{For a discussion of the potential cosmological
problems see for example \cite{Abel:1995wk}.}
it is possible
to write down a scale invariant superpotential%
\begin{equation}
 W= Y^{a
   b}_{e}\,\hat{\ell}_a^j\,\hat{e}_b^c\,\hat{H}_d^i\,\epsilon_{ij}+Y^{a
   b}_{d}\,\hat{q}^{j \alpha}_a\,\hat{d}_{\alpha
   b}^c\,\hat{H}_d^i\,\epsilon_{ij} + Y^{a b}_{u} \hat{q}^{i
   \alpha}_a\,\hat{u}_{\alpha b}^c\,\hat{H}_u^j\,\epsilon_{ij} +
 \lambda\,\hat{S}\,\hat{H}_u^i\,\hat{H}_d^j \epsilon_{ij} +
 \frac{1}{3}\kappa \,\hat{S}\,\hat{S}\,\hat{S}   \thickspace . \label{superpot}
\end{equation}
Here $a,b=1,2,3$ are generation indices and $i,j=1,2$ are $SU(2)_L$
gauge indices of the fundamental representation. $\epsilon_{ij}$ is
the totally anti-symmetric tensor. $Y_e,\,Y_d,\,Y_u$ are dimensionless
3x3 matrices of Yukawa couplings. The
soft SUSY breaking potential consists of masses for the scalar components of
the chiral superfields, gaugino mass terms as well as trilinear-scalar couplings:
\begin{eqnarray}
\nonumber \mathscr{V}_{\text{SB}} &=& m_{S}^2 |S|^2 + m_{H_u}^2 |H_u|^2 + m_{H_d}^2
|H_d|^2+ \tilde{q}^\dagger m_{\tilde{q}}^2 \tilde{q} +
\tilde{l}^\dagger m_{\tilde{l}}^2 \tilde{l} + \tilde{d}^\dagger
m_{\tilde{d}}^2 \tilde{d} + \tilde{u}^\dagger m_{\tilde{u}}^2
\tilde{u} \nonumber \\ && + \frac{1}{2}\left(M_1 \, \tilde{B}
\tilde{B} + M_2 \, \tilde{W}_i \tilde{W}^i + M_3 \, \tilde{g}_\alpha
\tilde{g}^\alpha + h.c.\right) \nonumber \\ && 
- H_u \tilde{q}
T_u\tilde{u}^\dagger +H_d \tilde{q} T_d \tilde{d}^\dagger + H_d
\tilde{l} T_e \tilde{e}^\dagger + T_\lambda S H_u H_d + \frac{1}{3} T_\kappa S S S \thickspace .
\end{eqnarray}
One of the appealing features of the NMSSM is that it solves the $\mu$
problem of the MSSM \cite{Kim:1983dt}: after SUSY breaking the scalar
component of $\hat{S}$, $S$, receives a vacuum expectation
value (VEV), denoted $v_s$, which leads to an effective mass term of
the Higgsinos
\begin{equation}
 \mu_{\rm{eff}} = \frac{1}{\sqrt{2}} \lambda v_s \, .
\end{equation}
Here, we have used the  decomposition
\begin{equation}
\label{VEVsinglet}
 S = \frac{1}{\sqrt{2}}  \left(\phi_s + i \sigma_s + v_s \right) \, .
\end{equation}
Since $v_s$ and thus also \(\mu_{\rm{eff}}\) are a consequence of SUSY breaking  
one finds that \(\mu_{\rm{eff}}\) is naturally of the order of
the SUSY breaking scale. 

$\phi_s$ and $\sigma_s$ mix together with the neutral Higgs boson of
the MSSM to form three CP even and two CP odd eigenstates, while the
fermionic component of $\hat{S}$ mixes after EWSB with the other four
neutralinos of the MSSM. In total, there are 29 mass eigenstates in
the NMSSM with $R$-parity $-1$, called sparticles: 12 squarks, 6 charged
sleptons, 3 neutral sleptons, 5 neutralinos, 2 charginos and 1 gluino.
With no \textit{a priori} model explaining the masses, these 29 states
lead to $29!\simeq8.8\cdot10^{30}$ possible mass orderings or
\textit{hierarchies}. Unfortunately, it is computationally impossible
to classify the dominant signatures of
this general setup. Therefore, we make the same
assumptions to reduce the number of hierarchies to a manageable amount
as for the MSSM in Ref.~\cite{Dreiner:2012wm}:
\begin{enumerate}
\item[(i)] The mixing between sparticles is sub-dominant, so we can identify
the mass eigenstates with the corresponding gauge eigenstates.  The
only exception are the Higgsinos, which we assume to be maximally
mixed.
\item[(ii)] The first and second generations of sfermions of the same
kind are degenerate in mass. We consider the third generation masses as
independent parameters, \textit{e.g.} for the sleptons
\begin{eqnarray}
m_{\tilde e L}&=& m_{\tilde \mu L}=m_{\tilde\nu_e}=m_{\tilde\nu_\mu} \equiv
m_{\tilde \ell,11} \\
m_{\tilde e R}&=& m_{\tilde \mu R}\equiv m_{\tilde e,11} \\
m_{\tilde \tau L}&=& m_{\tilde\nu_\tau}\equiv m_{\tilde \ell,33}\\
m_{\tilde\tau R}&\equiv& m_{\tilde e,33}\,,
\end{eqnarray}
and analogously for the squarks. 
\item[(iii)] The Higgsino mass mixing term is given by 
\begin{equation}
\mu_{\rm{eff}} = \frac{1}{\sqrt{2}} \lambda v_s \,,
\end{equation} 
and
the singlet mass is given by 
\begin{equation}
M_S = \frac{1}{\sqrt{2}} \kappa v_s \, .
\end{equation} 
\end{enumerate}
These three assumptions leave us with
15 relevant mass parameters,
\begin{eqnarray}
& M_1, M_2, M_3, \mu_{\rm{eff}}, M_S & \\ & m_{\tilde{e},11}, m_{\tilde{e},33},
  m_{\tilde{\ell},11}, m_{\tilde{\ell},33} & \\ 
& m_{\tilde{d},11},
  m_{\tilde{d},33}, m_{\tilde{u},11}, m_{\tilde{u},33},
  m_{\tilde{q},11}, m_{\tilde{q},33}\,, & 
\end{eqnarray}
and thus $15!$ different hierarchies.  Furthermore, the identification of the first and
second generation sfermions allows us to reduce the number of fields
we need to take into account in our analysis. We combine them because,
by assumption, they lead to the same signatures:
\begin{eqnarray}
(\tilde{e}_L/\tilde{\mu}_L)&\rightarrow&\tilde{\ell}, \nonumber\\
(\tilde{e}_R/\tilde{\mu}_R) &\rightarrow& \tilde{e}, \nonumber\\ 
(\tilde{d}_L/\tilde{s}_L/\tilde{u}_L/\tilde{c}_L) &\rightarrow&\tilde{q}, \nonumber\\
(\tilde{d}_R/\tilde{s}_R) &\rightarrow& \tilde{d}, \nonumber\\
(\tilde{u}_R/\tilde{c}_R)&\rightarrow& \tilde{u}, \nonumber \\
(\tilde{\nu}_e/\tilde{\nu}_\mu) &\rightarrow& \tilde{\nu},
\end{eqnarray} 
as well as the two Higgsino-like neutralinos. A
collection of the considered states as well as of the relevant mass
parameters is given in \TABLE~\ref{tab:fields}.

\begin{table}[!bt]
\begin{tabular}{|l|c|c|}
\hline
Particle & Name & Mass \\
\hline
Singlino-like neutralino & $\tilde{S}$  & $M_S$\\
Bino-like neutralino & $\tilde{B}$  & $M_1$\\
Wino-like neutralino & $\tilde{W}^0$ & $M_2$ \\
Higgsino-like neutralinos & $\tilde{H}^0$ & $\mu_{\rm{eff}}$ \\
Gluino & $\tilde{G}$ & $M_3$\\
Wino-like chargino & $\tilde{W}^\pm$ & $M_2$ \\
Higgsino-like chargino & $\tilde{H}^\pm$  & $\mu$\\
left-Squarks (1./2. generation) & $\tilde{q}_{1,2} \equiv \tilde{q}$ & $m_{\tilde{q},11}$  \\  
down-right Squarks (1./2. generation) & $\tilde{d},\tilde{s} \equiv \tilde{d}$& $m_{\tilde{d},11}$ \\ 
up-right Squarks (1./2. generation) & $\tilde{u}, \tilde{c} \equiv \tilde{u}$  & $m_{\tilde{u},11}$\\  
left charged sleptons (1./2. generation) & $\tilde{e}_L, \tilde{\mu}_L \equiv \tilde{l}$ & $m_{\tilde{l},11}$ \\
sneutrinos (1./2. generation) & $\tilde{\nu}_e, \tilde{\nu}_\mu \equiv \tilde{\nu}$ & $m_{\tilde{l},11}$ \\  
right sleptons (1./2. generation) & $\tilde{e}_R, \tilde{\mu}_R \equiv \tilde{e}$  & $m_{\tilde{e},11}$\\
left-Squarks (3. generation) & $\tilde{q}_3$ & $m_{\tilde{q},33}$ \\  
down-right Squarks (3. generation) & $\tilde{b}$ & $m_{\tilde{d},33}$\\ 
up-right Squarks (3. generation) & $\tilde{t}$ & $m_{\tilde{u},33}$\\  
left staus (3. generation) & $\tilde{\tau}_L$ & $m_{\tilde{l},33}$\\
sneutrinos (3. generation) & $\tilde{\nu}_\tau$ & $m_{\tilde{l},33}$\\  
right sleptons (3. generation) & $\tilde{\tau}_R$ & $m_{\tilde{e},33}$\\
\hline 
\end{tabular}
\caption{Particle content and relevant mass parameters.}
\label{tab:fields}
\end{table}

\section{Strategy for the analysis}
\label{sec:analysis}
We use for our analysis the same approach as for the MSSM in
Ref.~\cite{Dreiner:2012wm} which we summarize here. In total, we have $15! =
1.307.674.368.000\approx{ 1.3\cdot10^{12}}$ hierarchies.  Each one can
be denoted as a chain of fields in decreasing order of mass from left
to right:
\begin{equation}
 i_1 \dots i_n C r_1 \dots r_m
\end{equation}
$C$ denotes the lightest colored particle (LCP), excluding the third
generation. So $C$ is the lightest of the four fields $\tilde{G},\,
\tilde {q},\,\tilde{d}$ and $\tilde{u}$. The particles $\{i_k\}$ ($i$
for irrelevant) are all heavier, and contain among others the
remaining colored particles, other than possible third generation
squarks. The particles $\{r_k\}$ ($r$ for relevant) are all lighter
than $C$ and are potentially involved in the cascade decay of
$C$ and thus important for our analysis.  We assume that $C$
is the only directly produced particle at the LHC. We do not impose
any restrictions on the LSP, denoted $r_m$ above.

We are interested in the determination of the \textit{dominant decay
  chains} for all hierarchies. These are the decay chains $C \to r_i
\to \dots \to r_m = \text{LSP}$ that will dominantly happen at the LHC
for each hierarchy. Not all $r_j,\;j\in\{1,...,m\}$ are
  necessarily involved. In order to find them we apply the same
algorithm as in Refs.~\cite{Konar:2010bi,Dreiner:2012wm}:
\begin{enumerate}
 \item Find the SUSY particles which are lighter than the LCP and
 have the largest coupling to it.
 \item For each of these, search for the lighter
   particles with the largest coupling to it. The existing
   possibilities must be considered independently.
 \item Iterate step 2 until the LSP is reached.
\end{enumerate}
In principle, one can have more than one dominant decay chain for a
given hierarchy. That situation would correspond to decay chains with
similar rates at the LHC. Once the dominant decay chains are found, one
can determine their signature. These signatures, denoted here as
\textit{dominant signatures},\footnote{ If different signatures can be
  the result of a given decay chain we have chosen the one with the
  largest number of charged leptons.}  represent the main result of our study. 
  They are obtained by summing up the
decay products of all steps in the decay chain\footnote{See
  Ref.~\cite{Dreiner:2012wm} for
 the coupling strengths and the corresponding decay products in the MSSM, the NMSSM transitions are discussed below.}.

We
have considered as final state particles in our analysis
\begin{enumerate}
\item charged leptons ($l$), 
\item jets ($j$), 
\item massive
bosons ($v$) 
\item missing transverse energy ($\slashed{E}_T$) (neutrinos and 
neutralino and sneutrino LSP, for $R$pC).
\end{enumerate}
Note that \textit{massive bosons} stands for both gauge and Higgs
bosons including the scalar and pseudoscalar singlets.
\begin{table}
\centering
\begin{tabular}{|l c c|l c c|l c c|} 
\hline 
transition & strength & signature & transition &  strength & signature  & transition &  strength & signature \\ 
\hline
$\tilde{S}\leftrightarrow\tilde{H}^0$ & A & $v$&$\tilde{S}\leftrightarrow\tilde{H}^\pm$ & A & $v$&$\tilde{S}\leftrightarrow\tilde{l}$ & C & $v+l$\\ 
$\tilde{S}\leftrightarrow\tilde{d}$ & C & $j+v$&$\tilde{S}\leftrightarrow\tilde{q}$ & C & $j+v$&$\tilde{S}\leftrightarrow\tilde{u}$ & C& $j+v$\\ 
$\tilde{S}\leftrightarrow\tilde{W}^0$ & C & $2j+v$&$\tilde{S}\leftrightarrow\tilde{W}^\pm$ & C & $2j+v$&$\tilde{S}\leftrightarrow\tilde{e}$ & C& $v+l$\\ 
$\tilde{S}\leftrightarrow\tilde{\nu}$ & C& $l+v$&$\tilde{S}\leftrightarrow\tilde{t}$ & B& $j+v$&$\tilde{S}\leftrightarrow\tilde{b}$ & B& $j+v$\\ 
$\tilde{S}\leftrightarrow\tilde{q}_3$ & B& $j+v$&$\tilde{S}\leftrightarrow\tilde{\tau}_R$ & B& { $\slashed{E}_T+v$}&$\tilde{S}\leftrightarrow\tilde{\tau}_L$ &B& $j+v$\\ 
$\tilde{S}\leftrightarrow\tilde{\nu}_\tau$ &C& $l+v$&$\tilde{S}\leftrightarrow\tilde{B}$ &C& $2j+v$ & $\tilde{S}\leftrightarrow\tilde{G}$ &C& $2j+v$\\ 
\hline
\end{tabular}
\caption{Interactions of the singlino. We have considered for our
  analysis charged lepton ($l$), jets ($j$), massive bosons ($v$) and
  missing transversal energy ($\slashed{E}_T$) as signatures. The
  coupling strengths A, B and C depend on the value taken for
  $\lambda$, see text.}
\label{tab:signatures}
\end{table}
For the signatures and coupling strengths of the MSSM particle we refer to Ref.~\cite{Dreiner:2012wm}, where three 
different categories have been introduced to quantize the coupling strength
\begin{itemize}
 \item \textit{not suppressed}:  a two body decay mode which does not suffer from any mixing suppression
 \item \textit{suppressed}:  a two body decay mode which is mixing suppressed or a three body decay without additional suppression
 \item \textit{strongly suppressed}:  a three body decay with mixing suppression or a four body decay 
\end{itemize}
Note that some additional assumptions enter the definition of the
dominant signature for each transition. These have an impact on our
final results:
\begin{itemize}
\item We distinguish $\tilde{W}^0/\tilde{W}^\pm$,$\,\tilde{H}^0/\tilde
  {H}^\pm$, $\tilde{l}/\tilde{\nu}$ and $\tilde{\tau}/\tilde{\nu}_
  \tau$ in the decay chains because we differentiate between
  charged leptons and $\slashed{E}_T$ as a signature.
\item Emitted $\tau$'s are regarded as ordinary jets.
\item When, for a given transition, two different decay products with
  similar strengths are possible, we always choose the one with the
  largest amount of charged leptons. When the choice is between
  $\tau$s and $\slashed{E}_T$, we always choose $\slashed{E}_T$.
\item We disregard the possibility of degeneracies among fields of
  different types (with the exceptions mentioned above concerning
  first and second generation sfermions). Therefore, 2-body decays
  have no phase space suppression.
\item We do not treat jets originating from third generation quarks
  separately. 
\end{itemize}

For the singlino we distinguish three categories of couplings
(A,B,C) to the other SUSY particles, see
Table~\ref{tab:signatures}. The relative order between (A,B,C) and the
strengths relative to the MSSM transitions depend on the value taken
for $\lambda$. In the following we study three different cases:
\begin{enumerate}
 \item {\bf Small $\lambda$}: $\lambda$ is smaller than the electroweak gauge couplings. We can identify 
 \begin{itemize}
 \item A = suppressed 
 \item B = strongly suppressed
 \item C < strongly suppressed
 \end{itemize}
\item {\bf large $\lambda$}: $\lambda$ is not larger or smaller than
  the third generation Yukawa couplings.  This leads to the following
  size of singlino couplings:
 \begin{itemize}
 \item A = not suppressed
 \item B = suppressed 
 \item C = strongly suppressed
 \end{itemize}
 \item {\bf dominant $\lambda$}: $\lambda$ is larger than all other couplings. In that case we have 
 \begin{itemize}
\item A > not suppressed
\item not suppressed > B > suppressed
\item suppressed > C > strongly suppressed
\end{itemize}
\end{enumerate}

We exemplify this method with the hierarchy
\begin{equation}
i_1 \dots i_8 \tilde{G} \tilde{b} \tilde{H}^0 \tilde{S} \tilde{W}^0 \tilde{l} \tilde{B}
\end{equation}
For the first transition only one possibility exists
because the largest coupling is $\tilde{G}\rightarrow\tilde{b}$. 
For the second transition there are two dominant possibilities:
$\tilde{b}\rightarrow\tilde{H}^ 0,\,\tilde B$. Furthermore, the higgsino
couples with the same strength to the wino and to the bino, while the
coupling to the singlino depends on our choice of $\lambda$. If we
assume a dominant $\lambda$, the Higgsino will only decay into the
singlino. For large $\lambda$, the Higgsino decays with the same
probability into the wino, bino and singlino. Therefore, all three
branches have to be considered. For small $\lambda$, the Higgs decays
dominantly only to the Wino or Bino.

Moreover, the wino will always take the way via the
slepton to decay into the LSP. In contrast, as can be seen in \TABLE~\ref{tab:signatures} the singlino
couples to the wino, bino and slepton with the same strength. Hence, all three possibilities have to be
considered. In conclusion, depending on the choice of $\lambda$ different decay chains are possible: 
\begin{itemize}
 \item small $\lambda$: there are three dominant decay chains and two different dominant signatures:
\begin{align}
 \tilde{G} \rightarrow \tilde{b} \rightarrow 
\tilde{B} &: \hspace{0.3cm} 2 j \\
 \tilde{G} \rightarrow \tilde{b} \rightarrow \tilde{H}^0 \rightarrow 
\tilde{B} &: \hspace{0.3cm} 2 j + v\\
 \tilde{G} \rightarrow \tilde{b} \rightarrow \tilde{H}^0 \rightarrow 
\tilde{W}^0 \rightarrow \tilde{l} \rightarrow \tilde{B} &: \hspace{0.3cm}
  2 j + v+ 2 l
\end{align}
\item large $\lambda$: there are six dominant decay chains
\begin{align}
 \tilde{G} \rightarrow \tilde{b} \rightarrow 
\tilde{B} &: \hspace{0.3cm} 2 j \\
 \tilde{G} \rightarrow \tilde{b} \rightarrow \tilde{H}^0 \rightarrow \tilde{B} &: \hspace{0.3cm} 2 j + v\\
 \tilde{G} \rightarrow \tilde{b} \rightarrow \tilde{H}^0 \rightarrow \tilde{W}^0 \rightarrow \tilde{l} \rightarrow \tilde{B} &: \hspace{0.3cm}
  2 j + v+ 2 l \\
  \tilde{G} \rightarrow \tilde{b} \rightarrow \tilde{H}^0 \rightarrow \tilde{S} \rightarrow \tilde{B} &: \hspace{0.3cm} 4 j + 2 v\\
  \tilde{G} \rightarrow \tilde{b} \rightarrow \tilde{H}^0 \rightarrow \tilde{S} \rightarrow \tilde{l} \rightarrow  \tilde{B} &: \hspace{0.3cm} 2 j + 2 v +2 l\\
  \tilde{G} \rightarrow \tilde{b} \rightarrow \tilde{H}^0 \rightarrow \tilde{S} \rightarrow \tilde{W}^0 \rightarrow \tilde{l} \rightarrow  \tilde{B} &: \hspace{0.3cm} 4 j + 2 v + 2 l
\end{align}
\item dominant $\lambda$: there are four dominant decay chains
\begin{align}
\tilde{G} \rightarrow \tilde{b} \rightarrow 
\tilde{B} &: \hspace{0.3cm} 2 j \\
  \tilde{G} \rightarrow \tilde{b} \rightarrow \tilde{H}^0 \rightarrow \tilde{S} \rightarrow \tilde{B} &: \hspace{0.3cm} 4 j + 2 v\\
  \tilde{G} \rightarrow \tilde{b} \rightarrow \tilde{H}^0 \rightarrow \tilde{S} \rightarrow \tilde{l} \rightarrow  \tilde{B} &: \hspace{0.3cm} 2 j + 2 v + 2 l\\
  \tilde{G} \rightarrow \tilde{b} \rightarrow \tilde{H}^0 \rightarrow \tilde{S} \rightarrow \tilde{W}^0 \rightarrow \tilde{l} \rightarrow  \tilde{B} &: \hspace{0.3cm} 4 j + 2 v + 2 l
\end{align}

\end{itemize}

\section{Results}
\label{sec:results}

Here we present our results on whether a given signature appears in a
given setup or not. We have gone through the various sparticle
hierarchies and used only the dominant decay modes.  We have
categorized the signatures by the nature of the LSP: (a) neutral
($\tilde{S}$, $\tilde{B}$, $\tilde{W}^0$, $\tilde{H}^0$,
$\tilde{\nu}$, $\tilde{\nu}_\tau$), (b) charged ($\tilde{l}$,
$\tilde{e}$, $\tilde{ \tau}_L$, $\tilde{\tau}_R$, $\tilde{W}^+$,
$\tilde{H}^+$) and (c) colored ($\tilde {g}$, $\tilde{d}$,
$\tilde{u}$, $\tilde{q}$, $\tilde {b}_R$, $\tilde{t} _R$,
$\tilde{q}_3$).  As in our previous MSSM study~\cite{Dreiner:2012wm},
we have not restricted ourselves to the case of a neutral LSP which
could provide a valid dark matter candidate.  There are at least three
motivations to study also the case of a colored or charged LSP.  (i)
The relic density of the SUSY LSP could be so small to be
cosmologically negligible and dark matter is formed by other fields
like the axion
\cite{Weinberg:1977ma,Preskill:1982cy,Abbott:1982af,Dine:1982ah} or
the axino \cite{Rajagopal:1990yx,Covi:1999ty}. (ii)~There are
detailed experimental and theoretical collider studies for a charged
or a colored LSP in the literature
\cite{Abazov:2008qu,Aaltonen:2009kea,Raby:1997bpa,Baer:1998pg,Raby:1998xr,Dreiner:2008ca}.
(iii) Searches for $R$-hadrons have been performed at the Tevatron
\cite{Heister:2003hc,Arvanitaki:2005nq,Abazov:2007ht} and at the LHC
\cite{Aad:2011yf}.

For the R-parity conserving case we present our results in
  Tables \ref{tab:MSSM} -- \ref{tab:dominant}. We state in
each table whether a specific signature classified by the
number of charged leptons, jets and massive vector bosons appears
dominantly or not. Unlike in Ref.~\cite{Dreiner:2012wm}, we do
  not list the numerical frequency.  In the case of a colored or
charged LSP, we distinguish also between signatures without \MET\
(upper part of each cell) and with \MET\ (lower part of the cell). For
a neutral LSP like in the first part of \TABLE~\ref{tab:MSSM}, \MET\
is always present and we do not have to split the cells.

\subsection{MSSM}
\begin{table}[hbt]
\begin{tabular}{| c || c | c | c | c || c | c | c | c || c | c | c | c | } 
\hline 
& \multicolumn{4}{c||}{$n_v=0$}& \multicolumn{4}{c||}{$n_v=1$}& \multicolumn{4}{c|}{$n_v=2$}\\ \hline 
$n_j$ & $=0$ & $=1$ & $=2$ & $>2$ & $=0$ & $=1$ & $=2$ & $>2$ & $=0$ & $=1$ & $=2$ & $>2$  \\ \hline 
$n_l$ & \multicolumn{12}{c|}{neutral LSP} \\ \hline
0 & \tickNo & \tickYes & \tickYes & \tickYes & \tickNo & \tickYes & \tickYes & \tickYes &  \tickNo &\tickYes & \tickYes & \tickYes\\ 
\hline 
1 & \tickNo & \tickYes & \tickYes & \tickYes & \tickNo & \tickYes & \tickYes & \tickYes & \tickNo & \tickYes & \tickYes & \tickYes\\ 
\hline 
2 & \tickNo & \tickYes & \tickYes & \tickYes & \tickNo & \tickYes & \tickYes & \tickYes & \tickNo & \tickYes & \tickYes & \tickYes\\ 
\hline 
3 & \tickNo & \tickYes & \tickYes & \tickYes & \tickNo & \tickYes & \tickYes & \tickYes & \tickNo & \tickYes & \tickYes & \tickYes\\ 
\hline 
4 & \tickNo & \tickYes & \tickYes & \tickYes & \tickNo & \tickYes & \tickYes & \tickYes & \tickNo & \tickYes & \tickYes & \tickYes\\ 
\hline 
\hline 
$n_l$ & \multicolumn{12}{c|}{charged LSP} \\ \hline
\multirow{2}{*}{0} & \tickNo & \tickYes & \tickYes & \tickYes & \tickNo & \tickYes & \tickYes & \tickYes & \tickNo & \tickYes & \tickYes & \tickYes\\ 
                   & \tickNo & \tickYes & \tickYes & \tickYes & \tickNo & \tickYes & \tickYes & \tickYes & \tickNo & \tickYes & \tickYes & \tickYes\\ 
\hline 
\multirow{2}{*}{1} & \tickNo & \tickYes & \tickYes & \tickYes & \tickNo & \tickYes & \tickYes & \tickYes & \tickNo & \tickYes & \tickYes & \tickYes\\ 
                   & \tickNo & \tickYes & \tickYes & \tickYes & \tickNo & \tickYes & \tickYes & \tickYes & \tickNo & \tickYes & \tickYes & \tickYes\\ 
\hline 
\multirow{2}{*}{2} & \tickNo & \tickYes & \tickYes & \tickYes & \tickNo & \tickYes & \tickYes & \tickYes & \tickNo & \tickNo & \tickYes & \tickYes \\ 
                   & \tickNo & \tickYes & \tickYes & \tickYes & \tickNo & \tickYes & \tickYes & \tickYes & \tickNo & \tickYes & \tickYes & \tickYes\\ 
\hline 
\multirow{2}{*}{3} & \tickNo & \tickYes & \tickYes & \tickYes & \tickNo & \tickYes & \tickYes & \tickYes & \tickNo & \tickYes & \tickYes & \tickYes \\ 
                   & \tickNo & \tickYes & \tickYes & \tickYes & \tickNo & \tickYes & \tickYes & \tickYes & \tickNo & \tickYes & \tickYes & \tickYes \\ 
\hline 
\multirow{2}{*}{4} & \tickNo & \tickNo & \tickYes & \tickYes & \tickNo & \tickYes & \tickYes & \tickYes & \tickNo & \tickNo & \tickYes & \tickYes \\ 
                   & \tickNo & \tickNo & \tickYes & \tickYes & \tickNo & \tickYes & \tickYes & \tickYes & \tickNo & \tickNo & \tickYes & \tickYes \\ 
\hline 
\hline 
$n_l$ & \multicolumn{12}{c|}{colored LSP} \\ \hline
\multirow{2}{*}{0} & \tickYes & \tickYes & \tickYes & \tickYes & \tickNo & \tickNo & \tickYes & \tickYes & \tickNo & \tickNo & \tickYes & \tickYes \\ 
                   & \tickNo & \tickNo & \tickYes & \tickYes & \tickNo & \tickNo & \tickYes & \tickYes & \tickNo & \tickNo & \tickYes & \tickYes \\  
\hline 
\multirow{2}{*}{1} & \tickNo & \tickNo & \tickNo & \tickNo & \tickNo & \tickNo & \tickNo & \tickNo & \tickNo & \tickNo & \tickNo & \tickNo \\ 
                   & \tickNo & \tickNo & \tickYes & \tickYes & \tickNo & \tickNo & \tickYes & \tickYes & \tickNo & \tickNo & \tickYes & \tickYes \\ 
\hline 
\multirow{2}{*}{2} & \tickNo & \tickNo & \tickYes & \tickYes & \tickNo & \tickNo & \tickYes & \tickYes & \tickNo & \tickNo & \tickYes & \tickYes \\ 
                   & \tickNo & \tickNo & \tickYes & \tickYes & \tickNo & \tickNo & \tickYes & \tickYes & \tickNo & \tickNo & \tickYes & \tickYes \\ 
\hline 
\multirow{2}{*}{3} & \tickNo & \tickNo & \tickNo & \tickNo & \tickNo & \tickNo & \tickNo & \tickNo & \tickNo & \tickNo & \tickNo & \tickNo\\ 
                   & \tickNo & \tickNo & \tickYes & \tickYes & \tickNo & \tickNo & \tickYes & \tickYes & \tickNo & \tickNo & \tickYes & \tickYes\\ 
\hline 
\multirow{2}{*}{4} & \tickNo & \tickNo & \tickYes & \tickYes & \tickNo & \tickNo & \tickYes & \tickYes & \tickNo & \tickNo & \tickYes & \tickYes \\ 
                   & \tickNo & \tickNo & \tickYes & \tickYes & \tickNo & \tickNo & \tickYes & \tickYes & \tickNo & \tickNo & \tickYes & \tickYes \\ 
\hline 
\end{tabular} 
\caption{Results for the MSSM:  $n_v$
  denotes the number of bosons, $n_j$ the number of jets and $n_l$ the
  number of charged leptons from the single cascade chain. In the case
  of a charged and colored LSP, the upper entry in a given cell of the
  Table refers to no \MET\, the lower entry to \MET\ also being
  present. A neutral LSP is always counted as \MET. Signatures marked
  with \tickYes appear due to the dominant and best visible decay
  chains, while \tickNo\ could only be reached in subdominant
  cascades. }
\label{tab:MSSM}
\end{table}
Before we discuss the NMSSM, we recall in
\TABLE~\ref{tab:MSSM} all possible signatures for the MSSM given in
Ref.~\cite{Dreiner:2012wm}. The table has to be read as follows: we
separate the results in three big columns depending on the number of
massive bosons ($n_v$) in the final state. Each column
  is again divided into four smaller columns giving the number of
jets ($n_j$) appearing in the signatures. We distinguish the values
$n_j = 0,1,2$ and $n_j >2$. The rows show the number of charged
leptons ($n_l$). As already mentioned, while in the case of a neutral
LSP \MET\ is always present in each cascade, we have to distinguish
for a colored and charged LSP between events with and without
\MET. For that reason, the cells in the part of the table giving the
results for colored and charged are divided into an upper and lower
part. The upper gives the results for events without \MET, while the
lower one includes \MET.

We can see from \TABLE~\ref{tab:MSSM} that in the $R$-parity
conserving MSSM up to 2 massive vector bosons and up to 4 charged
leptons per cascade are possible. Furthermore, it is obvious that in
the case of a neutral or charged LSP at least one jet is emitted while
for a colored LSP also events without jets, charged leptons and bosons
are possible. This is the case when the LSP is the gluino or a squark
of the first two generations: in these scenarios the LSP is directly
dominantly produced at the LHC.

Interesting signatures for a neutral LSP are those with a large number
of charged leptons and a small number of jets. For instance, events
with $n_l = 4$, $n_v = 0$ and $n_j=1$ as well as a neutral LSP can be
the result of the decay chain
\begin{equation}
\label{eq:4jetEvent}
 \tilde{d} \to \tilde{e} \to \tilde{l} \to \tilde{W}^0 \, .
\end{equation}

 Recently, there have been some hints for events at
  the LHC with 7 jets and one lepton that cannot be explained by
  SM background \cite{Khoo}. This signature can be, for instance, a
  consequence of a bino LSP and the following hierarchy:
\begin{equation}
\tilde{G} \,  \tilde{q}_3 \, \tilde{W}^+ \, \tilde{\nu} \, \tilde{\tau}_L \, \tilde{H}^+ \, \tilde{t} \, \tilde{B}  \, .
\end{equation}
For this mass ordering, the following five cascades appear dominantly
\begin{align}
 \tilde{G}  \to \tilde{q}_3 \to  \tilde{B}&: \,\, 2j \\
 \tilde{G}  \to \tilde{q}_3 \to  \tilde H^+\to \tilde{B}&: \,\, 2j + v\\
 \tilde{G}  \to \tilde{q}_3 \to  \tilde H^+\to \tilde t\to \tilde{B}&: \,\, 4j \\
 \tilde{G}  \to \tilde{t} \to  \tilde{B}&: \,\, 2j \\
\tilde{G} \to  \tilde{q}_3 \to \tilde{W}^+ \to \tilde{\nu} \to \tilde{\tau}_R \to \tilde{H}^+ \to \tilde{t} \to \tilde{B} &: \,\, l + 5j\,.
\end{align}
Combining the first or the fourth cascade with the fifth can
explain the excess in this channel. See also  \cite{Lisanti:2011tm}.

In contrast, for a charged LSP, monojet events and four lepton tracks
are only possible if accompanied by exactly one massive
boson. Possible hierarchies for these signatures can easily be derived
from \EQ~(\ref{eq:4jetEvent}) by adding a stau or a charged Higgsino
to the end of the cascade:
\begin{align}
 \tilde{d} \to \tilde{e} \to \tilde{l} \to \tilde{W}^0  \to \tilde{\tau}_L \,\, &: \,\, 4 l + 2 j \\
 \tilde{d} \to \tilde{e} \to \tilde{l} \to \tilde{W}^0  \to \tilde{H}^+ \,\, &: \,\, 4 l +  j + v
\end{align}
Another important feature of a charged LSP is that it provides 
\textit{three}  signatures beside the four lepton monojet events which can
neither be reached by the other $R$pC cases nor by $R$pV scenarios:
$(n_v, n_j, n_l) = (0,1,2),$ $(1,1,0)$ and $(2,1,0)$.  Possible
cascades to obtain these signatures are 
\begin{align}
 (0,1,2) \, &: \, \, \tilde{d} \to \tilde{\nu} \to \tilde{W}^+ \label{chain1}\\
 (1,1,0) \, &: \, \, \tilde{q} \to \tilde{B} \to \tilde{H}^+ \\
 (2,1,0) \, &: \, \,  \tilde{q} \to \tilde{B} \to \tilde{H}^+ \to \tilde{W}^+ 
\end{align}

We summarize briefly the case of a colored LSP in the MSSM. The events
with one jet but nothing else are caused by a squark of the third
generation as the LSP and a produced gluino.  As soon as one lepton or
one massive boson is involved there have to be at least two jets: the
produced colored particle at the beginning of the cascade as well as
the stable colored particle at the end have to interact with
non-colored particles. Due to baryon number conservation at each
vertex at least two jets will appear. The reason that all events with
one or three charged leptons will also include neutrinos is, of
course, lepton number conservation.

We want to give here one example for a decay chain with the maximal
amount of charged leptons and massive bosons but the minimal amount of
jets: four leptons together with two massive bosons and two jets
follows from the decay chain
\begin{equation}
\label{eq:4jetColored}
\tilde{q} \to \tilde{B} \to \tilde{H}^0 \to \tilde{W}^0 \to \tilde{l} \to \tilde{e} \to \tilde{\tau}_R
\end{equation}

\subsection{NMSSM, Small $\lambda$}
\begin{table}[hbt]
\begin{tabular}{| c || c | c | c | c || c | c | c | c || c | c | c | c || c | c | c | c |} 
\hline 
& \multicolumn{4}{c||}{$n_v=0$}& \multicolumn{4}{c||}{$n_v=1$}& \multicolumn{4}{c||}{$n_v=2$}& \multicolumn{4}{c|}{$n_v=3$}\\ \hline 
$n_j$ & $=0$ & $=1$ & $=2$ & $>2$ & $=0$ & $=1$ & $=2$ & $>2$ & $=0$ & $=1$ & $=2$ & $>2$ & $=0$ & $=1$ & $=2$ & $>2$ \\ \hline 
$n_l$ & \multicolumn{16}{c|}{neutral LSP} \\ \hline
0 & \tickNo & \tickYes & \tickYes & \tickYes & \tickNo & \tickYes & \tickYes & \tickYes &  \tickNo &\tickYes & \tickYes & \tickYes &  \tickNo & \tickYes & \tickYes & \tickYes\\ 
\hline 
1 & \tickNo & \tickYes & \tickYes & \tickYes & \tickNo & \tickYes & \tickYes & \tickYes & \tickNo & \tickYes & \tickYes & \tickYes & \tickNo & \tickYes & \tickYes & \tickYes\\ 
\hline 
2 & \tickNo & \tickYes & \tickYes & \tickYes & \tickNo & \tickYes & \tickYes & \tickYes & \tickNo & \tickYes & \tickYes & \tickYes & \tickNo & \tickYes & \tickYes & \tickYes\\ 
\hline 
3 & \tickNo & \tickYes & \tickYes & \tickYes & \tickNo & \tickYes & \tickYes & \tickYes & \tickNo & \tickYes & \tickYes & \tickYes & \tickNo & \tickYes & \tickYes & \tickYes\\ 
\hline 
4 & \tickNo & \tickYes & \tickYes & \tickYes & \tickNo & \tickYes & \tickYes & \tickYes & \tickNo & \tickYes & \tickYes & \tickYes & \tickNo & \tickYes & \tickYes & \tickYes\\ 
\hline 
\hline 
$n_l$ & \multicolumn{16}{c|}{charged LSP} \\ \hline
\multirow{2}{*}{0} & \tickNo & \tickYes & \tickYes & \tickYes & \tickNo & \tickYes & \tickYes & \tickYes & \tickNo & \tickYes & \tickYes & \tickYes & \tickNo & \tickNo & \tickNo & \tickNo\\ 
 & \tickNo & \tickYes & \tickYes & \tickYes & \tickNo & \tickYes & \tickYes & \tickYes & \tickNo & \tickYes & \tickYes & \tickYes & \tickNo & \tickNo & \tickNo & \tickNo\\ 
\hline 
\multirow{2}{*}{1} & \tickNo & \tickYes & \tickYes & \tickYes & \tickNo & \tickYes & \tickYes & \tickYes & \tickNo & \tickYes & \tickYes & \tickYes & \tickNo & \tickYes & \tickYes & \tickYes\\ 
 & \tickNo & \tickYes & \tickYes & \tickYes & \tickNo & \tickYes & \tickYes & \tickYes & \tickNo & \tickYes & \tickYes & \tickYes & \tickNo & \tickYes & \tickYes & \tickYes\\ 
\hline 
\multirow{2}{*}{2} & \tickNo & \tickYes & \tickYes & \tickYes & \tickNo & \tickYes & \tickYes & \tickYes & \tickNo & \tickNo & \tickYes & \tickYes & \tickNo & \tickNo & \tickYes & \tickYes\\ 
 & \tickNo & \tickYes & \tickYes & \tickYes & \tickNo & \tickYes & \tickYes & \tickYes & \tickNo & \tickYes & \tickYes & \tickYes & \tickNo & \tickYes & \tickYes & \tickYes\\ 
\hline 
\multirow{2}{*}{3} & \tickNo & \tickYes & \tickYes & \tickYes & \tickNo & \tickYes & \tickYes & \tickYes & \tickNo & \tickYes & \tickYes & \tickYes & \tickNo & \tickYes & \tickYes & \tickYes\\ 
 & \tickNo & \tickYes & \tickYes & \tickYes & \tickNo & \tickYes & \tickYes & \tickYes & \tickNo & \tickYes & \tickYes & \tickYes & \tickNo & \tickYes & \tickYes & \tickYes\\ 
\hline 
\multirow{2}{*}{4} & \tickNo & \tickNo & \tickYes & \tickYes & \tickNo & \tickYes & \tickYes & \tickYes & \tickNo & \tickNo & \tickYes & \tickYes &  \tickNo &\tickNo & \tickYes & \tickYes\\ 
 & \tickNo & \tickNo & \tickYes & \tickYes & \tickNo & \tickYes & \tickYes & \tickYes & \tickNo & \tickNo & \tickYes & \tickYes & \tickNo & \tickNo & \tickNo & \tickNo\\ 
\hline 
\hline 
$n_l$ & \multicolumn{16}{c|}{colored LSP} \\ \hline
\multirow{2}{*}{0} & \tickYes & \tickYes & \tickYes & \tickYes & \tickNo & \tickNo & \tickYes & \tickYes & \tickNo & \tickNo & \tickYes & \tickYes & \tickNo & \tickNo & \tickNo & \tickNo\\ 
 & \tickNo & \tickNo & \tickYes & \tickYes & \tickNo & \tickNo & \tickYes & \tickYes & \tickNo & \tickNo & \tickYes & \tickYes & \tickNo & \tickNo & \tickNo & \tickNo\\ 
\hline 
\multirow{2}{*}{1} & \tickNo & \tickNo & \tickNo & \tickNo & \tickNo & \tickNo & \tickNo & \tickNo & \tickNo & \tickNo & \tickNo & \tickNo & \tickNo & \tickNo & \tickNo & \tickNo\\ 
 & \tickNo & \tickNo & \tickYes & \tickYes & \tickNo & \tickNo & \tickYes & \tickYes & \tickNo & \tickNo & \tickYes & \tickYes & \tickNo & \tickNo & \tickNo & \tickNo\\ 
\hline 
\multirow{2}{*}{2} & \tickNo & \tickNo & \tickYes & \tickYes & \tickNo & \tickNo & \tickYes & \tickYes & \tickNo & \tickNo & \tickYes & \tickYes & \tickNo & \tickNo & \tickNo & \tickNo\\ 
 & \tickNo & \tickNo & \tickYes & \tickYes & \tickNo & \tickNo & \tickYes & \tickYes & \tickNo & \tickNo & \tickYes & \tickYes & \tickNo & \tickNo & \tickNo & \tickNo\\ 
\hline 
\multirow{2}{*}{3} & \tickNo & \tickNo & \tickNo & \tickNo & \tickNo & \tickNo & \tickNo & \tickNo & \tickNo & \tickNo & \tickNo & \tickNo & \tickNo & \tickNo & \tickNo & \tickNo\\ 
 & \tickNo & \tickNo & \tickYes & \tickYes & \tickNo & \tickNo & \tickYes & \tickYes & \tickNo & \tickNo & \tickYes & \tickYes & \tickNo & \tickNo & \tickNo & \tickNo\\ 
\hline 
\multirow{2}{*}{4} & \tickNo & \tickNo & \tickYes & \tickYes & \tickNo & \tickNo & \tickYes & \tickYes & \tickNo & \tickNo & \tickYes & \tickYes & \tickNo & \tickNo & \tickNo & \tickNo\\ 
 & \tickNo & \tickNo & \tickYes & \tickYes & \tickNo & \tickNo & \tickYes & \tickYes & \tickNo & \tickNo & \tickYes & \tickYes & \tickNo & \tickNo & \tickNo & \tickNo\\ 
\hline 
\end{tabular} 
\caption{Results for the NMSSM assuming small $\lambda$. 
The notation is as in \TABLE~\ref{tab:MSSM}. In the case of a charged
and colored LSP, the upper entry in a given cell of the Table refers
to no \MET\, the lower entry to \MET\ also being present. A neutral
LSP is always counted as \MET.}
\label{tab:small}
\end{table}
We leave the MSSM and turn to the NMSSM. We start with the case of a
small coupling between the Higgs doublets and the gauge singlet. The
corresponding results are given in \TABLE~\ref{tab:small}. Since this
case is the MSSM limit of the NMSSM we do not expect large deviations
from the MSSM results presented in \TABLE~\ref{tab:MSSM}. This
assumption holds exactly for a colored LSP where there is no
difference between the NMSSM and the MSSM.  For a neutral and charged
LSP the MSSM and NMSSM agree in all possible signatures with $n_v <
3$. However, while it is not possible to get $n_v = 3$ in the MSSM, 
there are such events in the NMSSM. For instance, in the case of a
neutral LSP, $n_v = 3$, $n_j = n_l = 1$ can be a result of the decay
chain
\begin{equation}
 \tilde{q} \to \tilde{W}^0 \to \tilde{H}^0 \to \tilde{S} \to \tilde{\nu} \,
\end{equation}
while the same signature for a charged LSP appears in
\begin{equation}
 \tilde{q} \to \tilde{W}^0 \to \tilde{H}^0 \to \tilde{S} \to \tilde{e} \, . 
\end{equation}
One might wonder why $n_v=3$ events are not possible for a colored
LSP. The point is that a colored LSP which is not the LCP can only be
a third generation squark. These squarks couple, for small $\lambda$,
stronger to the Higgs fields than the singlet does. Therefore, the
Higgs decays directly to the LSP while for events with three bosons
the cascade $\tilde{H} \to \tilde{S} \to LCP$ is needed.

Another interesting observation is the fact that $n_v=3$ is only
possible for a charged LSP if at least one charged lepton is present,
while this is not the case for $n_v < 3$. To understand this, one must
know that all events with $n_l=0$ and $n_v=2$ have a charged Wino as
LSP. For instance, $n_j = 1$, $n_l=0$ and $n_v=2$ appears due to the
cascade
\begin{equation}
 \tilde{q} \to \tilde{B} \to \tilde{H}^0 \to \tilde{W}^+ \, .  
\end{equation}
To get a third boson $\tilde{S}$ must be present in the
cascade. However, $\tilde{H}^0$ couples stronger to the charged wino
than the singlino. Therefore, even if the the singlino is present the
Higgs decays dominantly to the LSP. Therefore, $n_v=3$ demands another
LSP with a weaker coupling to the Higgs fields. That is the case for
$\tilde{e}$ or $\tilde{l}$, and lepton number conservation explains
therefore the presence of at least on charged lepton track.

\subsection{NMSSM, Large $\lambda$}
\label{sec:large}
\begin{table}
\begin{tabular}{| c || c | c | c | c || c | c | c | c || c | c | c | c || c | c | c | c || c | c | c | c |} 
\hline 
& \multicolumn{4}{c||}{$n_v=0$}& \multicolumn{4}{c||}{$n_v=1$}& \multicolumn{4}{c||}{$n_v=2$}& \multicolumn{4}{c||}{$n_v=3$}& \multicolumn{4}{c|}{$n_v=4$}\\ \hline 
$n_j$ & $=0$ & $=1$ & $=2$ & $>2$ & $=0$ & $=1$ & $=2$ & $>2$ & $=0$ & $=1$ & $=2$ & $>2$ & $=0$ & $=1$ & $=2$ & $>2$ & $=0$ & $=1$ & $=2$ & $>2$ \\ \hline 
$n_l$ & \multicolumn{20}{c|}{neutral LSP} \\ \hline
0 & \tickNo & \tickYes & \tickYes & \tickYes &  \tickNo &\tickYes & \tickYes & \tickYes & \tickNo & \tickYes & \tickYes & \tickYes &\tickNo &  \tickYes & \tickYes & \tickYes & \tickNo & \tickNo & \tickYes & \tickYes\\ 
\hline 
1 & \tickNo & \tickYes & \tickYes & \tickYes & \tickNo & \tickYes & \tickYes & \tickYes & \tickNo & \tickYes & \tickYes & \tickYes & \tickNo & \tickYes & \tickYes & \tickYes & \tickNo & \tickNo & \tickYes & \tickYes\\ 
\hline 
2 & \tickNo & \tickYes & \tickYes & \tickYes & \tickNo & \tickYes & \tickYes & \tickYes & \tickNo & \tickYes & \tickYes & \tickYes & \tickNo & \tickYes & \tickYes & \tickYes & \tickNo & \tickNo & \tickYes & \tickYes\\ 
\hline 
3 & \tickNo & \tickYes & \tickYes & \tickYes & \tickNo & \tickYes & \tickYes & \tickYes & \tickNo & \tickYes & \tickYes & \tickYes & \tickNo & \tickYes & \tickYes & \tickYes & \tickNo & \tickNo & \tickYes & \tickYes\\ 
\hline 
4 & \tickNo & \tickYes & \tickYes & \tickYes & \tickNo & \tickYes & \tickYes & \tickYes & \tickNo & \tickYes & \tickYes & \tickYes & \tickNo & \tickYes & \tickYes & \tickYes & \tickNo & \tickNo & \tickYes & \tickYes\\ 
\hline 
\hline 
$n_l$ & \multicolumn{20}{c|}{charged LSP} \\ \hline
\multirow{2}{*}{0} & \tickNo & \tickYes & \tickYes & \tickYes & \tickNo & \tickYes & \tickYes & \tickYes & \tickNo & \tickYes & \tickYes & \tickYes & \tickNo & \tickNo & \tickNo & \tickYes & \tickNo & \tickNo & \tickNo & \tickYes\\ 
 & \tickNo & \tickYes & \tickYes & \tickYes & \tickNo & \tickYes & \tickYes & \tickYes & \tickNo & \tickYes & \tickYes & \tickYes & \tickNo & \tickYes & \tickYes & \tickYes & \tickNo & \tickNo & \tickNo & \tickYes\\ 
\hline 
\multirow{2}{*}{1} & \tickNo & \tickYes & \tickYes & \tickYes & \tickNo & \tickYes & \tickYes & \tickYes & \tickNo & \tickYes & \tickYes & \tickYes & \tickNo & \tickYes & \tickYes & \tickYes & \tickNo & \tickNo & \tickNo & \tickYes\\ 
 & \tickNo & \tickYes & \tickYes & \tickYes & \tickNo & \tickYes & \tickYes & \tickYes & \tickNo & \tickYes & \tickYes & \tickYes & \tickNo & \tickYes & \tickYes & \tickYes & \tickNo & \tickYes & \tickYes & \tickYes\\ 
\hline 
\multirow{2}{*}{2} & \tickNo & \tickYes & \tickYes & \tickYes & \tickNo & \tickYes & \tickYes & \tickYes & \tickNo & \tickYes & \tickYes & \tickYes &  \tickNo &\tickYes & \tickYes & \tickYes & \tickNo & \tickNo & \tickNo & \tickYes\\ 
 & \tickNo & \tickYes & \tickYes & \tickYes & \tickNo & \tickYes & \tickYes & \tickYes & \tickNo & \tickYes & \tickYes & \tickYes & \tickNo & \tickYes & \tickYes & \tickYes & \tickNo & \tickNo & \tickYes & \tickYes\\ 
\hline 
\multirow{2}{*}{3} & \tickNo & \tickYes & \tickYes & \tickYes &  \tickNo & \tickYes & \tickYes & \tickYes & \tickNo & \tickYes & \tickYes & \tickYes & \tickNo & \tickYes & \tickYes & \tickYes & \tickNo & \tickNo & \tickNo & \tickYes\\ 
 & \tickNo & \tickYes & \tickYes & \tickYes & \tickNo & \tickYes & \tickYes & \tickYes & \tickNo & \tickYes & \tickYes & \tickYes & \tickNo & \tickYes & \tickYes & \tickYes & \tickNo & \tickNo & \tickYes & \tickYes\\ 
\hline 
\multirow{2}{*}{4} & \tickNo & \tickNo & \tickYes & \tickYes &\tickNo & \tickYes & \tickYes & \tickYes & \tickNo & \tickYes & \tickYes & \tickYes & \tickNo & \tickNo & \tickNo & \tickYes & \tickNo & \tickNo & \tickNo & \tickYes\\ 
 & \tickNo & \tickNo & \tickYes & \tickYes & \tickNo & \tickYes & \tickYes & \tickYes & \tickNo & \tickYes & \tickYes & \tickYes & \tickNo & \tickYes & \tickYes & \tickYes & \tickNo & \tickNo & \tickNo & \tickYes\\ 
\hline 
\multirow{2}{*}{5} & \tickNo & \tickNo & \tickNo & \tickNo &  \tickNo & \tickNo & \tickNo & \tickNo & \tickNo & \tickNo & \tickYes & \tickNo & \tickNo & \tickNo & \tickNo & \tickNo & \tickNo & \tickNo & \tickNo & \tickNo\\ 
                   & \tickNo & \tickNo & \tickNo & \tickNo & \tickNo & \tickNo & \tickNo & \tickNo & \tickNo & \tickNo & \tickNo & \tickNo & \tickNo & \tickNo & \tickNo & \tickNo  & \tickNo & \tickNo & \tickNo & \tickNo\\ 
\hline 
\hline 
$n_l$ & \multicolumn{20}{c|}{colored LSP} \\ \hline
\multirow{2}{*}{0} & \tickYes & \tickYes & \tickYes & \tickYes & \tickNo & \tickNo & \tickYes & \tickYes & \tickNo & \tickNo & \tickYes & \tickYes & \tickNo & \tickNo & \tickYes & \tickYes & \tickNo & \tickNo & \tickNo & \tickYes\\ 
 & \tickNo & \tickNo & \tickYes & \tickYes & \tickNo & \tickNo & \tickYes & \tickYes & \tickNo & \tickNo & \tickYes & \tickYes & \tickNo & \tickNo & \tickYes & \tickYes & \tickNo & \tickNo & \tickYes & \tickYes\\ 
\hline 
\multirow{2}{*}{1} & \tickNo & \tickNo & \tickNo & \tickNo & \tickNo & \tickNo & \tickNo & \tickNo & \tickNo & \tickNo & \tickNo & \tickNo & \tickNo & \tickNo & \tickNo & \tickNo & \tickNo & \tickNo & \tickNo & \tickNo\\ 
 & \tickNo & \tickNo & \tickYes & \tickYes & \tickNo & \tickNo & \tickYes & \tickYes & \tickNo & \tickNo & \tickYes & \tickYes & \tickNo & \tickNo & \tickYes & \tickYes & \tickNo & \tickNo & \tickNo & \tickYes\\ 
\hline 
\multirow{2}{*}{2} & \tickNo & \tickNo & \tickYes & \tickYes & \tickNo & \tickNo & \tickYes & \tickYes & \tickNo & \tickNo & \tickYes & \tickYes & \tickNo & \tickNo & \tickYes & \tickYes & \tickNo & \tickNo & \tickNo & \tickYes\\ 
 & \tickNo & \tickNo & \tickYes & \tickYes & \tickNo & \tickNo & \tickYes & \tickYes & \tickNo & \tickNo & \tickYes & \tickYes & \tickNo & \tickNo & \tickYes & \tickYes & \tickNo & \tickNo & \tickNo & \tickYes\\ 
\hline 
\multirow{2}{*}{3} & \tickNo & \tickNo & \tickNo & \tickNo & \tickNo & \tickNo & \tickNo & \tickNo & \tickNo & \tickNo & \tickNo & \tickNo & \tickNo & \tickNo & \tickNo & \tickNo & \tickNo & \tickNo & \tickNo & \tickNo\\ 
 & \tickNo & \tickNo & \tickYes & \tickYes & \tickNo & \tickNo & \tickYes & \tickYes & \tickNo & \tickNo & \tickYes & \tickYes & \tickNo & \tickNo & \tickYes & \tickYes & \tickNo & \tickNo & \tickNo & \tickYes\\ 
\hline 
\multirow{2}{*}{4} & \tickNo & \tickNo & \tickYes & \tickYes & \tickNo & \tickNo & \tickYes & \tickYes & \tickNo & \tickNo & \tickYes & \tickYes & \tickNo & \tickNo & \tickYes & \tickYes & \tickNo & \tickNo & \tickNo & \tickYes\\ 
 & \tickNo & \tickNo & \tickYes & \tickYes & \tickNo & \tickNo & \tickYes & \tickYes & \tickNo & \tickNo & \tickYes & \tickYes & \tickNo & \tickNo & \tickYes & \tickYes & \tickNo & \tickNo & \tickNo & \tickYes\\ 
\hline 
\end{tabular} 
\caption{Results for the NMSSM assuming large $\lambda$.  The notation
  is as in \TABLE~\ref{tab:MSSM}. In the case of a charged and colored
  LSP, the upper entry in a given cell of the Table refers to no
  \MET\, the lower entry to \MET\ also being present. A neutral LSP is
  always counted as \MET.}
\label{tab:large}
\end{table}

As a next scenario we assume that $\lambda$ is comparable to
the third generation Yukawa couplings. In this case, the Higgsinos
decay dominantly into the singlet and the third generation squarks
with the same probability if both are lighter than $\mu_{eff}$. The
resulting dominant signatures are given in \TABLE~\ref{tab:large}. The
main result is that signatures with up to four massive bosons are
possible. This observation is independent of the nature of the LSP.

For a neutral LSP the dominant signatures with $n_v < 3$ agree
completely with the MSSM results, while this is not the case for a
charged or colored LSP. We will explain this below. First, some words
about the neutral LSP: the NMSSM with large $\lambda$ and a neutral
LSP can produce all signatures dominantly with $n_j \geqslant 2$, $n_v
\leq 4$ and $n_l \leq 4$. One important result for the $n_v = 4$
signatures is that they can only be obtained by the transition $LCP
\to \tilde{B} \to \tilde{H} \to \tilde{W} \to \tilde{S} \to LSP$ or
$LCP \to \tilde{W} \to \tilde{H} \to \tilde{B} \to \tilde{S} \to LSP$,
\textit{i.e.}  $\tilde{W}$ and $\tilde{B}$ have to be heavier than the
singlino and therefore the LSP must be a neutral slepton.

Finally, we want to point out that there is also one signature for a
neutral LSP and large $\lambda$ which can not be reached by
any other configuration in the NMSSM: $(n_v,n_j, n_l) = (4,2,4)$ (with
\MET). Let us give a possible hierarchy which leads to the
corresponding cascade:
\begin{align}
 (4,2,4) : & \hspace{0.5cm} \tilde{q} \to \tilde{W} \to \tilde{H} \to \tilde{B} \to \tilde{e} \to \tilde{S} \to \tilde{\tau}_R \to \tilde{l} \to \tilde{\nu}_\tau 
\end{align}

The case of a charged LSP is even more interesting. Not only are
signatures with $n_v > 2$ present, while they were impossible in the
MSSM, but also for $n_v \leq 2$ a new signature with five charged
lepton tracks arises. In the MSSM there was an upper limit of
four. The signature $(n_v,n_j, n_l) = (2,2,5)$ without \MET\ can be
for instance produced via the cascade:
\begin{align}
\label{ref:LargeCharged56a}
 (2,2,5)  : & \hspace{0.5cm} \tilde{G} \to \tilde{S} \to \tilde{l} \to \tilde{W}^0 \to \tilde{B} \to \tilde{e}  
\end{align}
But even for less than four charged leptons there are signatures with
$n_v = 2$ which do not appear dominantly in the MSSM with a charged
LSP, but which are present in the NMSSM: $(n_v,n_j, n_l) =
(2,1,2),(2,1,4),(2,1,4)+\slashed{E}_T$. Cascades resulting in these
signatures are the following:
\begin{align}
 (2,1,2)  : & \hspace{0.5cm} \tilde{q} \to \tilde{H}^0 \to \tilde{S} \to \tilde{\nu} \to \tilde{W}^+  \\
 (2,1,4) : & \hspace{0.5cm}  \tilde{q} \to \tilde{W}^0 \to \tilde{l} \to \tilde{B} \to \tilde{e} \to \tilde{S} \to \tilde{H}^+ \\
 (2,1,4) + \slashed{E}_T : & \hspace{0.5cm} \tilde{q} \to \tilde{H}^{0}
 \to \tilde{S} \to \tilde{\nu} \to \tilde{W}^+ \to \tilde{B} \to 
\tilde{e} 
\end{align}
The only remaining case for large $\lambda$ is the one with a colored
LSP. The results are presented in the last part of
\TABLE~\ref{tab:large}. Also in this setup up to four massive bosons
are possible. However, each of them is accompanied by at least two
jets. The reason is the same as for the MSSM: there are at least two
colored vertices and baryon number is conserved. As in the scenario
with a neutral LSP, all signatures with $n_v < 3$ agree exactly with
those of the MSSM.  Furthermore, all possible dominantly appearing
signatures with $n_v \geqslant 3$ can also be obtained in the case of
a neutral or charged LSP. Hence, it is difficult to find a smoking gun
signature for the NMSSM with large $\lambda$ and a colored LSP.

\subsection{NMSSM, Dominant $\lambda$}
\begin{table}
\begin{tabular}{| c || c | c | c | c || c | c | c | c || c | c | c | c || c | c | c | c |} 
\hline 
& \multicolumn{4}{c||}{$n_v=0$}& \multicolumn{4}{c||}{$n_v=1$}& \multicolumn{4}{c||}{$n_v=2$}& \multicolumn{4}{c|}{$n_v=3$}\\ \hline 
$n_j$ & $=0$ & $=1$ & $=2$ & $>2$ & $=0$ & $=1$ & $=2$ & $>2$ & $=0$ & $=1$ & $=2$ & $>2$ & $=0$ & $=1$ & $=2$ & $>2$ \\ \hline 
\hline
 $n_l$ & \multicolumn{16}{c|}{neutral LSP} \\ \hline
0 & \tickNo &\tickYes & \tickYes & \tickYes & \tickNo &\tickYes & \tickYes & \tickYes &\tickNo & \tickYes & \tickYes & \tickYes & \tickNo & \tickYes & \tickYes & \tickYes\\ 
\hline 
1 & \tickNo &\tickYes & \tickYes & \tickYes & \tickNo &\tickYes & \tickYes & \tickYes & \tickNo &\tickYes & \tickYes & \tickYes & \tickNo & \tickYes & \tickYes & \tickYes\\ 
\hline 
2 & \tickNo &\tickYes & \tickYes & \tickYes &\tickNo & \tickYes & \tickYes & \tickYes &\tickNo & \tickYes & \tickYes & \tickYes & \tickNo & \tickYes & \tickYes & \tickYes\\ 
\hline 
3 & \tickNo &\tickYes & \tickYes & \tickYes &\tickNo & \tickYes & \tickYes & \tickYes & \tickNo &\tickYes & \tickYes & \tickYes &\tickNo &  \tickYes & \tickYes & \tickYes\\ 
\hline 
4 &\tickNo & \tickYes & \tickYes & \tickYes &\tickNo & \tickYes & \tickYes & \tickYes & \tickNo & \tickYes & \tickYes & \tickYes &\tickNo & \tickYes & \tickYes & \tickYes\\ 
\hline 
\hline 
 $n_l$ & \multicolumn{16}{c|}{charged LSP} \\ \hline
\multirow{2}{*}{0} & \tickNo & \tickYes & \tickYes & \tickYes & \tickNo & \tickYes & \tickYes & \tickYes & \tickNo & \tickYes & \tickYes & \tickYes & \tickNo & \tickNo & \tickNo & \tickYes\\ 
 & \tickNo & \tickYes & \tickYes & \tickYes & \tickNo & \tickYes & \tickYes & \tickYes & \tickNo & \tickYes & \tickYes & \tickYes & \tickNo & \tickYes & \tickYes & \tickYes\\ 
\hline 
\multirow{2}{*}{1} & \tickNo & \tickYes & \tickYes & \tickYes & \tickNo & \tickYes & \tickYes & \tickYes & \tickNo & \tickYes & \tickYes & \tickYes & \tickNo & \tickYes & \tickYes & \tickYes\\ 
 & \tickNo & \tickYes & \tickYes & \tickYes & \tickNo & \tickYes & \tickYes & \tickYes & \tickNo & \tickYes & \tickYes & \tickYes & \tickNo & \tickYes & \tickYes & \tickYes\\ 
\hline 
\multirow{2}{*}{2} & \tickNo & \tickYes & \tickYes & \tickYes & \tickNo & \tickYes & \tickYes & \tickYes & \tickNo & \tickYes & \tickYes & \tickYes &  \tickNo & \tickYes & \tickYes & \tickYes\\ 
 & \tickNo & \tickYes & \tickYes & \tickYes & \tickNo & \tickYes & \tickYes & \tickYes & \tickNo & \tickYes & \tickYes & \tickYes & \tickNo & \tickYes & \tickYes & \tickYes\\ 
\hline 
\multirow{2}{*}{3} & \tickNo & \tickYes & \tickYes & \tickYes & \tickNo & \tickYes & \tickYes & \tickYes & \tickNo & \tickYes & \tickYes & \tickYes & \tickNo & \tickYes & \tickYes & \tickYes\\ 
 &  \tickNo & \tickYes & \tickYes & \tickYes & \tickNo & \tickYes & \tickYes & \tickYes & \tickNo & \tickYes & \tickYes & \tickNo & \tickYes & \tickYes & \tickYes & \tickYes\\ 
\hline 
\multirow{2}{*}{4} & \tickNo & \tickNo & \tickYes & \tickYes & \tickNo & \tickYes & \tickYes & \tickYes & \tickNo & \tickYes & \tickYes & \tickYes & \tickNo & \tickNo & \tickNo & \tickYes\\ 
 &\tickNo & \tickNo & \tickYes & \tickYes & \tickNo & \tickYes & \tickYes & \tickYes & \tickNo & \tickYes & \tickYes & \tickYes & \tickNo & \tickYes & \tickYes & \tickYes\\ 
\hline 
\multirow{2}{*}{5} & \tickNo & \tickNo & \tickNo & \tickNo & \tickNo & \tickNo & \tickNo & \tickNo & \tickNo & \tickNo & \tickYes & \tickNo & \tickNo & \tickNo & \tickNo & \tickNo\\ 
 & \tickNo & \tickNo & \tickNo & \tickNo & \tickNo & \tickNo & \tickNo & \tickNo & \tickNo & \tickNo & \tickNo & \tickNo &  \tickNo & \tickNo & \tickNo & \tickNo\\ 
\hline 
\hline 
 $n_l$ & \multicolumn{16}{c|}{colored LSP} \\ \hline
\multirow{2}{*}{0} & \tickYes & \tickYes & \tickYes & \tickYes & \tickNo & \tickNo & \tickYes & \tickYes & \tickNo & \tickNo & \tickYes & \tickYes & \tickNo & \tickNo & \tickYes & \tickYes\\ 
 & \tickNo & \tickNo & \tickYes & \tickYes & \tickNo & \tickNo & \tickYes & \tickYes & \tickNo & \tickNo & \tickYes & \tickYes & \tickNo & \tickNo & \tickYes & \tickYes\\ 
\hline 
\multirow{2}{*}{1} & \tickNo & \tickNo & \tickNo & \tickNo & \tickNo & \tickNo & \tickNo & \tickNo & \tickNo & \tickNo & \tickNo & \tickNo & \tickNo & \tickNo & \tickNo & \tickNo\\ 
 & \tickNo & \tickNo & \tickYes & \tickYes & \tickNo & \tickNo & \tickYes & \tickYes & \tickNo & \tickNo & \tickYes & \tickYes & \tickNo & \tickNo & \tickYes & \tickYes\\ 
\hline 
\multirow{2}{*}{2} & \tickNo & \tickNo & \tickYes & \tickYes & \tickNo & \tickNo & \tickYes & \tickYes & \tickNo & \tickNo & \tickYes & \tickYes & \tickNo & \tickNo & \tickYes & \tickYes\\ 
 & \tickNo & \tickNo & \tickYes & \tickYes & \tickNo & \tickNo & \tickYes & \tickYes & \tickNo & \tickNo & \tickYes & \tickYes & \tickNo & \tickNo & \tickYes & \tickYes\\ 
\hline 
\multirow{2}{*}{3} & \tickNo & \tickNo & \tickNo & \tickNo & \tickNo & \tickNo & \tickNo & \tickNo & \tickNo & \tickNo & \tickNo & \tickNo & \tickNo & \tickNo & \tickNo & \tickNo\\ 
 & \tickNo & \tickNo & \tickYes & \tickYes & \tickNo & \tickNo & \tickYes & \tickYes & \tickNo & \tickNo & \tickYes & \tickYes & \tickNo & \tickNo & \tickYes & \tickYes\\ 
\hline 
\multirow{2}{*}{4} & \tickNo & \tickNo & \tickYes & \tickYes & \tickNo & \tickNo & \tickYes & \tickYes & \tickNo & \tickNo & \tickYes & \tickYes & \tickNo & \tickNo & \tickYes & \tickYes\\ 
 & \tickNo & \tickNo & \tickYes & \tickYes & \tickNo & \tickNo & \tickYes & \tickYes & \tickNo & \tickNo & \tickYes & \tickYes & \tickNo & \tickNo & \tickYes & \tickYes\\ 
\hline 
\end{tabular} 
\caption{Results for the NMSSM assuming dominant $\lambda$.  The
  notation is as in \TABLE~\ref{tab:MSSM}. In the case of a charged
  and colored LSP, the upper entry in a given cell of the Table refers
  to no \MET\, the lower entry to \MET\ also being present. A neutral
  LSP is always counted as \MET.}
\label{tab:dominant}
\end{table}
If one drops the assumption that $\lambda$ is perturbative up to the
GUT scale and assumes instead a $\lambda$SUSY scenario
\cite{Barbieri:2006bg}, it is possible that $\lambda$ is even much
larger than the top Yukawa coupling. We now discuss this case.
The corresponding results are given in \TABLE~\ref{tab:dominant}. It
can be seen that only signatures with at most three massive bosons
show up dominantly. This is a bit surprising because for large
$\lambda$ four bosons have been possible. However, we have already
seen that the case of $n_v =4$ demands the transitions $\tilde{B} \to
\tilde{H} \to \tilde{W} \to (\dots \to) \tilde{S}$ or $\tilde{W} \to
\tilde{H} \to \tilde{B} \to (\dots \to) \tilde{S}$. These transitions
are highly suppressed for dominant $\lambda$ because the Higgsino will
decay prominently directly to the singlino.

Comparing \TABLE~\ref{tab:dominant} with \TABLE~\ref{tab:large} it
turns out that exactly the same signatures with $n_v < 4$ appear in
the case of a large and of a dominant $\lambda$. That means that there
is no unique setup which would be a strong indication for the NMSSM
with a dominant $\lambda$ coupling between the singlet and the
Higgs fields.

\medskip

To summarize the main results for the possible signatures in the
NMSSM: we have seen that for a small value of $\lambda$ the case of a
colored LSP is exactly as in the MSSM, while for neutral and charged
LSPs signatures with $n_v = 3$ are possible. The upper limit on
charged lepton tracks in this scenario is four. In
contrast, for a large or dominant $\lambda$ up to 5 charged leptons
can be emitted during the cascade decays. However, this happens
dominantly only for $n_v = 2$. While it is possible to get $n_v = 3$
for all three kinds of LSPs for a dominant $\lambda$, for a large
singlino coupling even $n_v = 4 $ is possible. Finally, comparing
\TABLE~\ref{tab:small} to
\ref{tab:dominant}, one can see that events with $n_l=4$, $n_j=1$ and
without a massive boson ($n_v = 0$) are only possible for a neutral
LSP independent of the assumed order of $\lambda$.

\subsection{NMSSM with $R$-parity violation}
We have also derived all possible signatures in the case of $R$-parity
violation
\cite{Hall:1983id,Dreiner:1997uz,Allanach:2003eb,Bhattacharyya:1997vv,Barger:1989rk,Allanach:1999ic,Hirsch:2000ef,Barbier:2004ez}. For
this purpose we used the dominant decay modes already presented in
Ref.~\cite{Dreiner:2012wm}. While a detailed discussion of the data is
beyond the scope of this paper, we want to point out some unique
signatures\footnote{A pdf file with the tables of all possible
  $R$pV-NMSSM signatures can be obtained by email from the
  authors.}\!\!. The four possible types of couplings in the general,
$R$-parity violating NMSSM are the same as for the MSSM\footnote{The
  singlet superfield $\hat S$ allows for additional $R$-parity
  violating terms. This is for example the case of $\hat S\hat{\ell}_i
  \hat{H}_u $, as in the $\mu\nu$SSM
  \cite{LopezFogliani:2005yw}. After electroweak symmetry breaking
  this operator leads to the NMSSM with an effective bilinear
  term. Therefore, the collider phenomenology of the $\mu\nu$SSM
  cannot be distinguished from that of the NMSSM with an explicit
  $\epsilon_i \hat{\ell}_i \hat{H}_u$ superpotential term
  \cite{Ghosh:2008yh,Bartl:2009an}. Similarly, the bilinear term can
  also be generated in models with spontaneous $R$-parity violation
  \cite{Masiero:1990uj} where, in contrast, the phenomenology can be
  altered due to the presence of a majoron \cite{Hirsch:2008ur}.} and
read
\begin{equation} \label{rpv-superpot}
W_{\slashed{R}} = \epsilon_i \hat{\ell}_i \hat{H}_u +\frac{1}{2}  
\lambda_{ijk} \hat{\ell}_i \hat{\ell}_j \hat{e}_k^c  + \frac{1}{2}  
\lambda^{'}_{ijk} \hat{q}_i \hat{d}_j^c \hat{\ell}_k + \frac{1}{2} 
\lambda^{''}_{ijk} \hat{u}_i^c \hat{d}_j^c \hat{d}_k^c \,.
\end{equation}
In the following we assume that only one of these couplings is present
at once. In addition, we assume that only the LSP decays through an
$R$pV operator. The decay modes are listed in
Table~\ref{tab:RpVmodes}.

\begin{table}
  \begin{tabular}{|c|cccc|}
   \hline 
                   & $\epsilon$ & $\lambda$ & $\lambda'$ & $\lambda''$ \\
 \hline
 $\tilde{B}$ & $h^0 \nu$ & $l^+ l^- \nu$ & $l^\pm q \bar{q}'$ & $q q' q''$ \\
 $\tilde{W}^\pm$ & $Z^0 l^\pm$ & $3 l^\pm$ & $l^\pm q \bar{q}$ & $q q' q''$ \\
 $\tilde{W}^0$ & $W^\pm l^\mp$ & $l^+ l^- \nu$ & $l^\pm q \bar{q}'$ & $q q' q''$ \\
 $\tilde{G}$ & $q \bar{q}' l^\pm$ & $q \bar{q} l^+ l^- \nu$ & $l^\pm q \bar{q}'$ & $q q' q''$ \\
 $\tilde{H}^\pm$ & $Z^0 l^\pm$ & $3 l^\pm$ & $l^\pm q \bar{q}$ & $q q' q''$ \\
 $\tilde{H}^0$ & $W^\pm l^\mp$ &$l^+ l^- \nu$ & $l^\pm q \bar{q}'$ & $q q' q''$ \\
 $\tilde{q}$ & $l^\pm q$ & $q l^+ l^- \nu$ & $l^\pm q$ & $4 q$ \\
 $\tilde{d}$ & $l^\pm q$ & $q l^+ l^- \nu$ & $l^\pm q$ & $q q'$ \\
 $\tilde{u}$ & $q \nu$ & $q l^+ l^- \nu$ & $l^\pm q \bar{q}' q''$ & $q q'$ \\
$\tilde{l}$ & $q \bar{q}'$ & $l^\pm \nu$ & $q \bar{q}'$ & $q q' q'' l^\pm$ \\
$\tilde{\nu}$ & $q \bar{q}$ & $l^+ l^-$ & $q \bar{q}$ & $q q' q'' \nu$ \\
$\tilde{e}$ & $l^\pm \nu$ & $l^\pm \nu$ & $l^\pm l^\pm q \bar{q}'$ & $q q' q'' l^\pm$ \\
$\tilde{q}_3$ & $l^\pm q$ & $q l^+ l^- \nu$ & $l^\pm q$ & $4 q$ \\
$\tilde{b}_R$ & $q \nu$ & $q l^+ l^- \nu$ & $q \nu$ & $q q'$ \\
$\tilde{t}_R$ & $l^\pm q$ & $q l^+ l^- \nu$ & $l^\pm q \bar{q}' q''$ & $q q'$ \\
$\tilde{\tau}_L$ & $q \bar{q}'$ & $l^\pm \nu$ & $q \bar{q}'$ & $q q' q'' \tau$ \\
$\tilde{\nu}_\tau$  & $q \bar{q}$ & $l^+ l^-$ & $q \bar{q}$ & $q q' q'' \nu$ \\
$\tilde{\tau}_R$ & $\tau \nu$ & $l^\pm \nu$ & $l^\pm \nu q \bar{q}$ & $q q' q'' \tau$ \\
$\tilde{S}$ & $W^\pm l^\mp$ &$l^+ l^- \nu$ & $l^\pm q \bar{q}'$ & $q q' q''$ \\
\hline
\end{tabular}
\caption{Dominant $R$-parity violating decay modes of the LSP
\cite{Dreiner:1997uz,Porod:2000hv,Hirsch:2003fe,Barbier:2004ez,Dreiner:1991pe}. Note that
we have chosen charged lepton final states over \MET\ and thus neglected the
decay $\tilde B\rightarrow\nu q\bar q'$, for example.}
\label{tab:RpVmodes}
\end{table}

\subsubsection{$\epsilon_i \hat{\ell}_i \hat{H}_u$}
In the MSSM with bilinear $R$pV at most 5 charged leptons and two
massive scalars are possible in a cascade. In contrast, 
in the NMSSM it is
possible to get six or seven leptons for a large
or dominant $\lambda$. Six leptons are also possible for the
$R$pC NMSSM but only without \MET. Including the $\epsilon$ term, we
can get them also with
\MET: 
\begin{align}
 (2,>2,6) + \slashed{E}_T : &  \hspace{0.5cm}  \hspace{0.5cm}  
\tilde{G} \to \tilde{e} \to \tilde{H}^0 \to \tilde{S} \to \tilde{l} 
\to \tilde{W}^0 \to \tilde{B} \to \tilde{\tau}_R \to \tau \, + \, \nu \,,
\end{align}
where the $\tau$ in the final state is counted as an additional
jet. For small and dominant $\lambda$ also signatures with $n_v = 4$
are possible. This is interesting because $n_v=4$ without $R$pV is
only possible for large $\lambda$, as we have seen. The reason is that
the $R$pV decays of a wino, bino or Higgsino LSP produce additional
bosons. On the other hand, $n_v=5$ is not possible for large $\lambda$
despite what one might expect. The point is that $n_v = 4$ in the $R$pC
case is only possible if $\tilde{B}$, $\tilde{W}$ and $\tilde{H}$ are
heavier than the singlino as we have discussed in
sec.~\ref{sec:large}.  Therefore, they cannot be the LSP and their
$R$pV decay modes play only a sub-leading role.

\subsubsection{$\lambda_{ijk} \hat{\ell}_i \hat{\ell}_j \hat{e}_k^c$}
The lepton number violating interaction $\hat{\ell}_i \hat{\ell}_j
\hat{e}_k^c$ can cause up to seven leptons in a cascade, but only two
bosons in the MSSM. In contrast, in the NMSSM extended by
  this operator, it is possible to obtain up to four massive bosons
and seven leptons. Four bosons are only possible for large $\lambda$
\begin{align}
 (4,1,6) + \slashed{E}_T : &  \hspace{0.5cm}  \tilde{q} \to \tilde{B} \to \tilde{H} \to \tilde{W}^0 \to \tilde{l} \to \tilde{S} \to \tilde{e} \to \tilde{\nu}_\tau \to l l
\end{align}
while three bosons are emitted dominantly for all $\lambda$'s. A signature not dominantly arising in the $R$pC case but present here for all values of $\lambda$ is the one with six lepton tracks: 
\begin{align}
 (3,1,6) + \slashed{E}_T : &  \hspace{0.5cm}  \tilde{q} \to \tilde{B} \to \tilde{H}^0 \to \tilde{W}^0 \to \tilde{l} \to \tilde{e} \to \tilde{S} \to \nu ll
\end{align}
The case of seven leptons also exists for all $\lambda$'s and can occur for large $\lambda$ as a result of
\begin{align}
 (2,>2,7) + \slashed{E}_T : &  \hspace{0.5cm}   \tilde{G} \to \tilde{t} \to \tilde{H}^0 \to \tilde{S} \to \tilde{e} \to \tilde{B} \to \tilde{W}^0 \to \tilde{\nu} \to \tilde{\nu}_\tau \to l l   
\end{align}

\subsubsection{$\lambda^{'}_{ijk} \hat{q}_i \hat{d}_j^c \hat{\ell}_k$}
The operator $\hat{q}_i \hat{d}_j^c \hat{\ell}_k$ produces in general
additional jets. In the MSSM the number of charged leptons is
restricted to at most five, and of massive bosons to two. In contrast,
we can find in the NMSSM with the same operator
\begin{align}
 (2,>2,7)  \hspace{0.3cm} (\text{large, dominant}\, \lambda) : & \hspace{0.5cm}  \tilde{G} \to \tilde{S} \to \tilde{l} \to \tilde{W}^0 \to \tilde{B} \to \tilde{e} \to l l q \bar{q}'   \\
 (3,>2,5) + \slashed{E}_T \hspace{0.3cm} (\text{all}\, \lambda) : &  \hspace{0.5cm}   \tilde{q} \to \tilde{t} \to \tilde{H}^0 \to \tilde{W}^0 \to \tilde{l} \to \tilde{S} \to \tilde{e} \to \tilde{\tau}_R \to l  \nu q \bar{q}  \\  
 (4,>2,5) + \slashed{E}_T \hspace{0.3cm} (\text{large}\, \lambda) : & \hspace{0.5cm}  \tilde{q} \to \tilde{B} \to \tilde{H}^0 \to \tilde{W}^0 \to \tilde{l} \to \tilde{S} \to \tilde{e} \to \tilde{\tau}_R \to l  \nu q \bar{q}    
\end{align}

\subsubsection{$\lambda^{''}_{ijk} \hat{u}_i^c \hat{d}_j^c \hat{d}_k^c$}
Cascades involving $\lambda^{''}_{ijk} \hat{u}_i^c \hat{d}_j^c
\hat{d}_k^c$ can produce even more jets than in the case of
$\lambda^{'}_{ijk} \hat{q}_i \hat{d}_j^c \hat{\ell}_k$, but the number
of charged leptons and massive bosons in the MSSM are limited as in
the $R$-parity conserving case: only $n_l \leq 4$ and
$n_v leq 2$ is possible. If we go to the NMSSM, we can find for
all possible values of $\lambda$ also signatures with $n_v=3$, while
for large $\lambda$ also $n_v = 4$ is possible. However, the
signatures are the same as for $R$pC, except for the additional
jets. The same holds for signatures with six charged leptons, which do
not appear for small $\lambda$, but in the other two cases the same
results as for $R$-parity conservation are obtained.

\section{Conclusion}
\label{sec:conclusion}
We have discussed in this paper the collider signatures dominantly appearing 
in a very general realization of the NMSSM.  It is based on 15 unrelated mass 
parameters which lead to $15! \simeq 1.3\cdot 10^{12}$ possible particle mass 
orderings. We have studied three possible scenarios for the singlet-Higgs 
coupling $\lambda$. We checked possible signatures to discriminate these 
three scenarios among each other but also from the MSSM. For small $\lambda$, 
the signatures for all LSPs are identical to the MSSM as long as less than 3 massive 
bosons are present. Signatures with 3 bosons are not possible in the MSSM, 
but can appear in the NMSSM for all possible ranges of $\lambda$. Furthermore, 
in the case of a large but not dominant $\lambda$, even up to four massive 
bosons can be emitted during the cascade decays. This is also the only difference 
between the case of a large and dominant $\lambda$: all signatures with less 
bosons are identical. On the other side, for both setups, signatures with less than 
three massive bosons arise and these do not appear dominantly in the MSSM. For 
instance, in the MSSM there can be at most four charged lepton tracks if $R$-parity 
is conserved, while we find also hierarchies in the NMSSM which can dominantly 
emit five charged leptons.

We pointed out a hierarchy which can dominantly lead to seven jets and
one lepton, which could explain the observed excess at the LHC.

We briefly commented on the NMSSM with $R$-parity violation and the 
possible signatures. We found that, depending on the $R$-parity violating 
parameters, outstanding signatures with up to five massive bosons or 
seven charged leptons are possible.

\section*{Acknowledgements}
We thank Werner Porod for discussions and collaboration in the early
stage of this project. AV acknowledges support from the ANR project
CPV-LFV-LHC {NT09-508531}. HKD acknowledges support from
BMBF grant 00160200.

\end{document}